\documentclass[a4paper,10pt,twoside]{cpc-hepnp}

\usepackage{multicol}
\usepackage{graphicx}
\usepackage{booktabs}
\usepackage{amssymb,bm,mathrsfs,bbm,amscd}
\usepackage[tbtags]{amsmath}
\usepackage{lastpage}

\def\Im{\mathop{\rm Im}\nolimits}

\begin{document}

\fancyhead[co]{\footnotesize D. Fern\'andez-Fraile \& A. G\'omez Nicola: Transport coefficients of a massive pion gas}

\footnotetext[0]{Received 15 December 2009}

\title{Transport coefficients of a massive pion gas}

\author{
D. Fern\'andez-Fraile $^{1;1)}$\email{danfer@th.physik.uni-frankfurt.de}
\quad A. G\'omez Nicola $^{2;2)}$\email{gomez@fis.ucm.es}
}
\maketitle

\address{%
1~(Institut f\"ur Theoretische Physik, Johann Wolfgang Goethe-Universit\"at, Max-von-Laue-Str. 1, 60438 Frankfurt am Main, Germany)\\
2~(Departamento de F\'isica Te\'orica II, Universidad Complutense, 28040 Madrid, Spain)\\
}

\begin{abstract}
We review or main results concerning the transport coefficients of a light meson gas, in particular we focus on the case of a massive pion
gas. Leading order results according to the chiral power-counting are presented for the DC electrical conductivity, thermal conductivity,
shear viscosity, and bulk viscosity. We also comment on the possible
correlation between the bulk viscosity and the trace anomaly in QCD,
as well as the relation between unitarity and a minimum of the
quotient $\eta/s$ near the phase transition.
\end{abstract}



\begin{multicols}{2}

%%%%%%%%%%%%%%%%%%%%%%%%%%%%%%%
\section{Introduction}
%%%%%%%%%%%%%%%%%%%%%%%%%%%%%%%

The calculation of transport coefficients in quantum field theory at
intermediate and strong coupling is still a challenge from both the
analytical and the numerical points of view. Due to their intrinsic non-perturbative
nature, even in weakly interacting theories a resummation of an
infinite number of diagrams is needed in order to obtain the
leading-order result. In the
strongly coupled regime, the only reliable method available is the
AdS/CFT correspondence \cite{Son:07}, although it is only applicable to a limited
class of field theories. On the other hand, lattice techniques for
extracting these quantities are still in their infancy, and the
inclusion of a finite quark chemical potential in these calculations
makes of it an even more difficult challenge because of the sign problem \cite{Aarts:09}.

Transport coefficients are essential inputs to describe the evolution of systems not far from
equilibrium. In particular, during the last years there has been a very
active effort to analyze them from both the theoretical and phenomenological points of
view in the context of heavy-ion collisions, condensed matter physics,
astrophysics and cosmology. We will be here interested in QCD
at zero baryon chemical potential and temperatures below the chiral
phase transition, where the hadrons are the relevant degrees of
freedom. Near $T_\mathrm{c}$ the QCD coupling constant is not small, and usual
perturbative techniques give in principle not reliable results
there. We will present results concerning a diagrammatic calculation
of transport coefficients in the context of Chiral Perturbation Theory
for a pion gas, which is the dominant component of the hadron gas at low
temperatures, and we will extrapolate these results up to temperatures
close to $T_\mathrm{c}$ by using unitarization methods.  

%%%%%%%%%%%%%%%%%%%%%%%%%%%%%%%%%%%
\section{Transport coefficients}
%%%%%%%%%%%%%%%%%%%%%%%%%%%%%%%%%%%

Consider a quantum field characterized by two conserved currents, the
energy-momentum tensor $T_{\mu\nu}$ and another current $N_\mu$. In
the local rest frame of the quantum fluid, we can derive two constitutive relations
by considering a small departure from a reference equilibrium state:
\begin{align}
&\langle \hat{T}_{ij}\rangle=P_\mathrm{eq}\delta_{ij}+\eta\left(\partial_iU_j+\partial_jU_i+\frac{2}{3}\delta_{ij}\partial_kU^k\right)-\zeta\delta_{ij}\partial_kU^k\ ,\label{shearbulk}\\
&\langle\hat{T}^{i0}\rangle-h\langle\hat{N}^i\rangle=\kappa\frac{T^2}{h}\partial_i\left(\frac{\mu}{T}\right)\ ,\label{thermal}
\end{align}
where $\langle\cdot\rangle$ denotes a non-equilibrium statistical average, $U^\mu$ is the fluid velocity normalized as
$U^2=1$, $P_\mathrm{eq}\equiv\langle-\hat{T}^k_k/3\rangle_\mathrm{eq}$ is the (equilibrium)
pressure, which satisfies an
equation of state $P_\mathrm{eq}=P_\mathrm{eq}(\epsilon,n)$ in terms of the energy density
$\epsilon\equiv\langle\hat{T}_{00}\rangle$ and the charge density
$n\equiv\langle \hat{N}^0\rangle$, and $h=(\epsilon+P)/n$ is the
enthalpy. The shear viscosity $\eta$,
the bulk viscosity $\zeta$, and the thermal conductivity $\kappa$,
determine the departure from the equilibrium state. On the other hand,
if we consider an electrically charged field, the DC conductivity
$\sigma$ is defined through the constitutive relation $\langle \hat{J}^i\rangle=\sigma
E^i$, where $\langle \hat{J}^i\rangle$ is the electric current produced after applying a
(classical) constant electric field $E^i$ to the system.

From (\ref{shearbulk}) we see that the shear viscosity is related to departures
from equilibrium due to transverse flow of momentum, whereas the bulk
viscosity is related to variations of pressure, energy density, and
charge density due to uniform
compression or expansion of the system, because $\Delta P-c_\epsilon\Delta\epsilon
-c_n\Delta n=-\zeta\nabla\cdot\bm{U}$, where $\Delta
P\equiv\langle-T^k_k/3\rangle-P_\mathrm{eq}(\epsilon_0,n_0)$,
$c_\epsilon\equiv (\partial P_\mathrm{eq}/\partial
\epsilon)(\epsilon_0,n_0)$, $c_n\equiv (\partial P_\mathrm{eq}/\partial n)(\epsilon_0,n_0)$, $\Delta\epsilon\equiv
\epsilon-\epsilon_0$, and $\Delta n\equiv n-n_0$, with $\epsilon_0$ and $n_0$ the values
corresponding to the reference equilibrium state. And from
(\ref{thermal}) we see that the thermal conductivity is identically
zero if there is not a conserved current in the system besides the
energy-momentum tensor (in the case of the pion gas we consider that
the conserved charge is the total number of pions \cite{FernandezFraile:09}).

\vspace{0.3cm}In Linear Response Theory (LRT), a transport coefficient
$\mathcal{T}$ is expressed in terms of \emph{retarded} correlation functions through
the Kubo formulas:
\begin{align}
&\mathcal{T}=C_\mathcal{T}\lim_{\omega\rightarrow
  0^+}\lim_{p\rightarrow
  0^+}\frac{\partial\rho_\mathcal{T}(\omega,p)}{\partial\omega}\ ,\label{LRT}\\
&\rho_\mathcal{T}(\omega,p)=2 \Im \mathrm{i}\int\mathrm{d}^4x\
\mathrm{e}^{\mathrm{i}p\cdot x}\theta(t)\langle[\hat{\mathcal{S}}_\mathcal{T}^\alpha(x),\hat{\mathcal{S}}^\mathcal{T}_\alpha(0)]\rangle_\mathrm{eq}\ ,
\end{align}
where $\alpha$ represents some set of indices. For
the different transport coefficients considered here
\cite{FernandezFraile:09}\footnote{There is typo in the formula (3.43) of
  Ref. \cite{FernandezFraile:09}, where $c_s^2$ and $\mu$ should be
  substituted by $c_\epsilon$ and $c_n$ respectively.}:
\begin{align}
&C_\sigma\equiv-\frac{1}{6}\ ,\ C_\eta\equiv\frac{1}{20}\ ,\ C_\zeta\equiv\frac{1}{2}\ ,\ C_\kappa\equiv-\frac{1}{6T}\ ,\notag\\
&\hat{\mathcal{S}}_\sigma^i\equiv\hat{J}^i\ ,\notag\\
&\hat{\mathcal{S}}_\eta^{ij}\equiv\hat{T}^{ij}-g^{ij}\hat{T}^k_k/3\ ,\notag\\
&\hat{\mathcal{S}}_\zeta\equiv-\hat{T}^k_k/3-c_\epsilon\hat{T}^{00}-c_n\hat{N}^0\ ,\notag\\
&\hat{\mathcal{S}}_\kappa^i\equiv\hat{T}^{0i}-h\hat{N}^i\ .\notag
\end{align}
%%%%%%%%%%%%%%%%%%%%%%%
\subsection{Diagrams vs. kinetic theory}
%%%%%%%%%%%%%%%%%%%%%%%
Due to the zero momentum limit, the leading-order contribution (in
powers of the
coupling constant) to transport coefficients obtained by using the
formula (\ref{LRT}) involves the resummation of an infinite number of
\emph{ladder} diagrams, as was first shown by Jeon and Yaffe\cite{Jeon:95,Jeon:96} in the case of a
$\lambda\phi^4$ theory. This is essentially because every two
propagators which share the same four-momentum in that limit give a
contribution $\propto 1/\lambda^2$. In the case of the bulk viscosity,
in addition, another set of diagrams called \emph{chain}  (or \emph{bubble})
diagrams give an important vertex correction that has to be taken
into account. In those papers it was also proved that, in the
weakly coupled regime, this resummation is equivalent to solving the
Boltzmann equation corresponding to an effective kinetic theory where
thermal effects are incorporated into the dispersion relation of the
quasi-particles as well as into their scattering
amplitudes. Consequently, it is in principle more convenient
technically to adopt
the effective kinetic theory approach to calculate transport coefficients in
weakly coupled theories \cite{Arnold:00,Arnold:06}. However, in the
context of ChPT where we have carried out our analysis for the pion gas, the relevant power-counting for estimating the
contribution from the different Feynman diagrams suggests that the
leading order of transport coefficients is given by a one-loop diagram
with dressed pion propagators \cite{FernandezFraile:09,FernandezFraile:06}, which
simplifies significantly the calculation.

%%%%%%%%%%%%%%%%%%%%%%%
\subsection{Results for a pion gas}
%%%%%%%%%%%%%%%%%%%%%%%
We dress the thermal pion propagators by considering the lowest order
process according to ChPT in pion-pion scattering, and we unitarize
the corresponding amplitude by the Inverse Amplitude Method (IAM)
\cite{FernandezFraile:09,FernandezFraile:07,Cabrera:09} (the ChPT
expansion eventually violates unitarity as we increase the collision energy). Unitarization
allows to extend the range of applicability of the ChPT expansion to
higher temperatures, and correctly reproduces the resonances
$f_0(600)$ and $\rho(770)$ in the pion-pion scattering channels
$(I,J)=(0,0),(1,1)$ respectively. This unitarization also allows the
study of the evolution of these resonances with temperature and baryon
density, as was presented in the references \cite{FernandezFraile:07,Cabrera:09}. 

In Fig. \ref{plotstransport} we can see that unitarization changes the
qualitative behavior of transport coefficients with temperature. In
particular it makes the quotient $\eta/s$, with $s$ the entropy
density, to satisfy the conjectured lower bound $1/4\pi$ from AdS/CFT \cite{Son:07},
and the value near $T_\mathrm{c}$ is close to the full hadron
resonance + Hagedorn states gas result \cite{NoronhaHostler:09}. In the case of
the bulk viscosity, unitarization enhances the anomalous peak close to
$T_\mathrm{c}$ \cite{FernandezFraile:09-2}, nevertheless the quotient
$\zeta/s$ is still smaller than $\eta/s$, although vertex corrections
in the case of $\zeta$ may play an important role \cite{Jeon:95,Chen:09}. 

\end{multicols}
\ruleup
\begin{center}
\includegraphics[height=3.3cm]{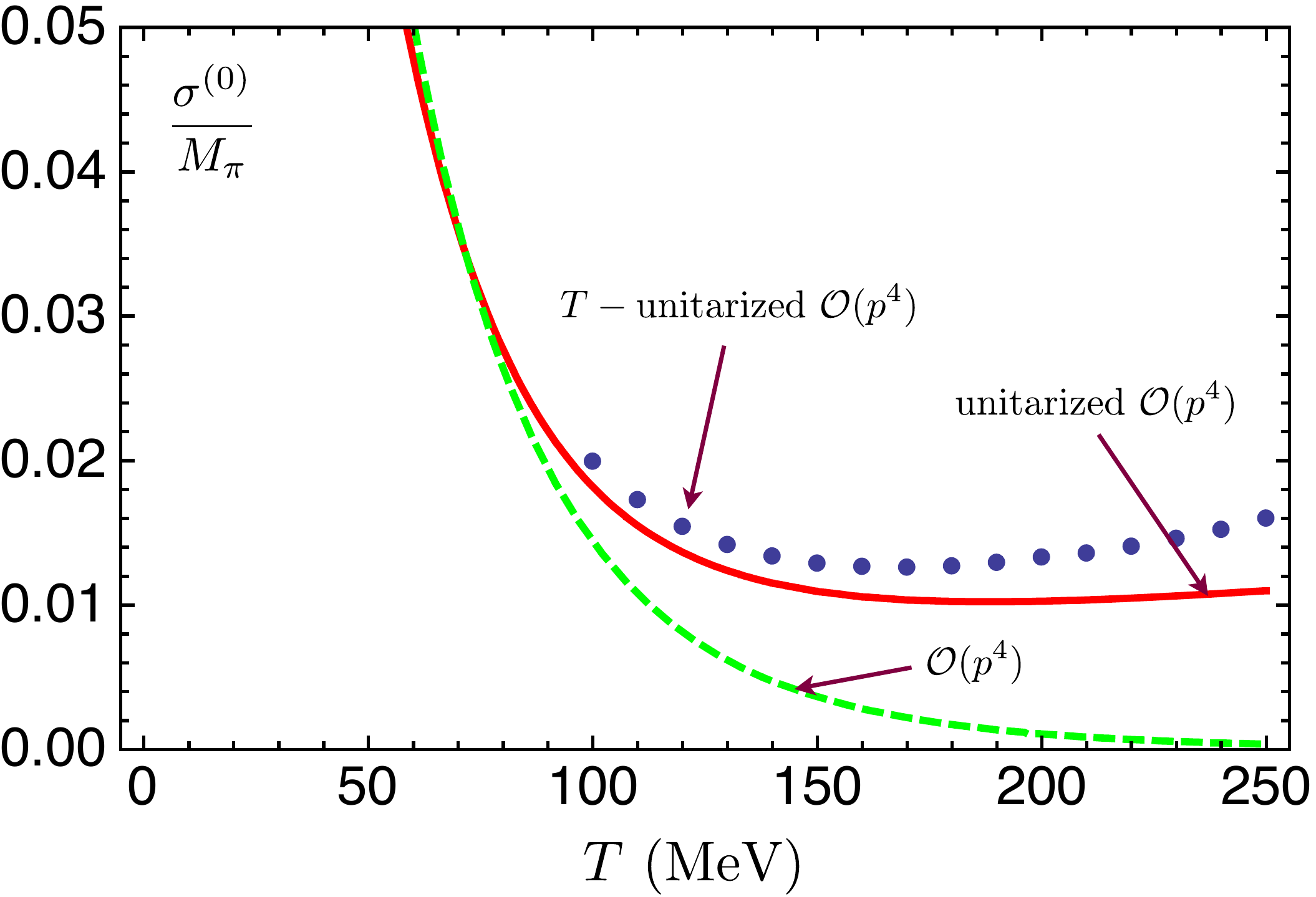}\hspace{0.5cm}\includegraphics[height=3.3cm]{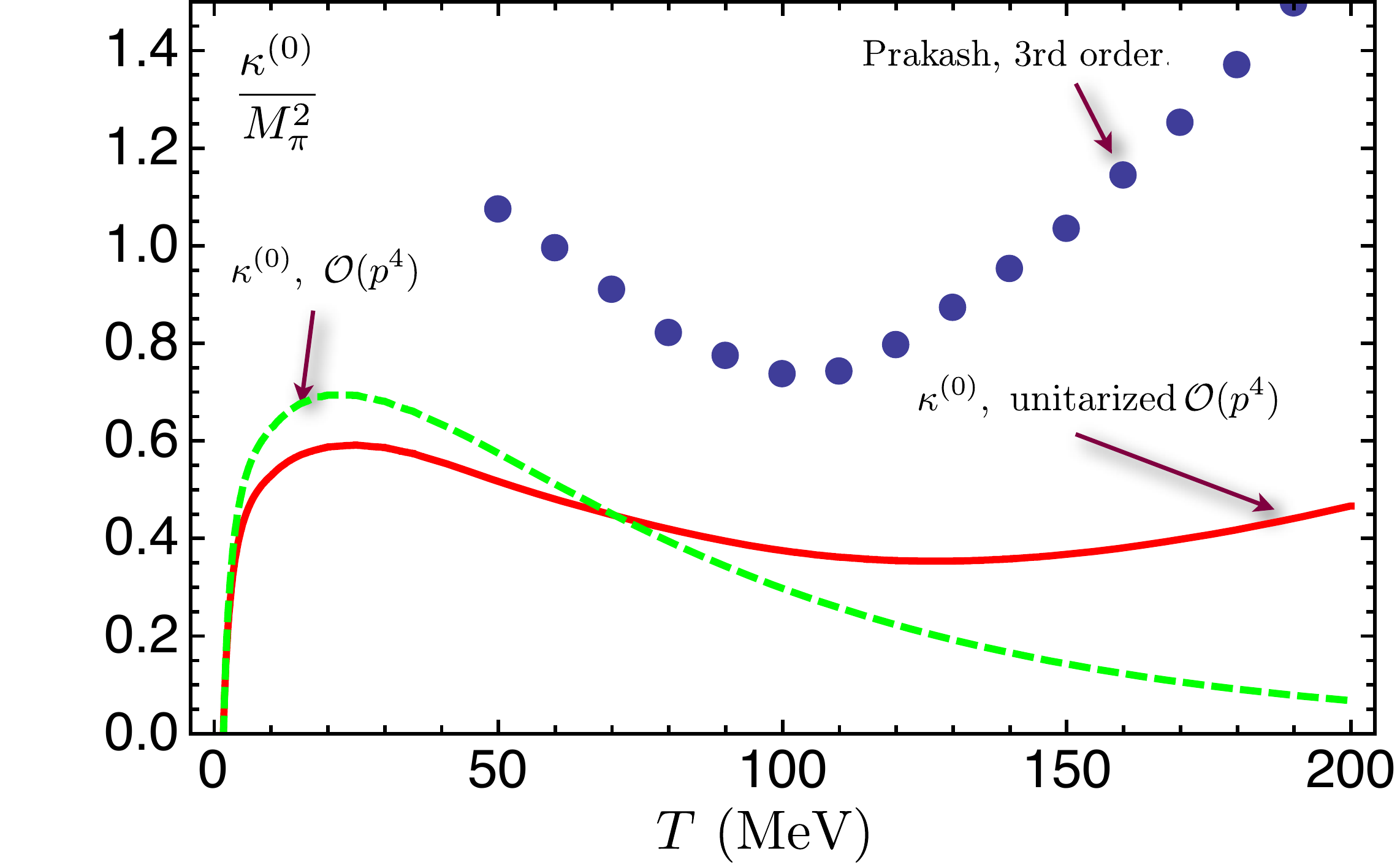}\hspace{0.5cm}\includegraphics[height=3.3cm]{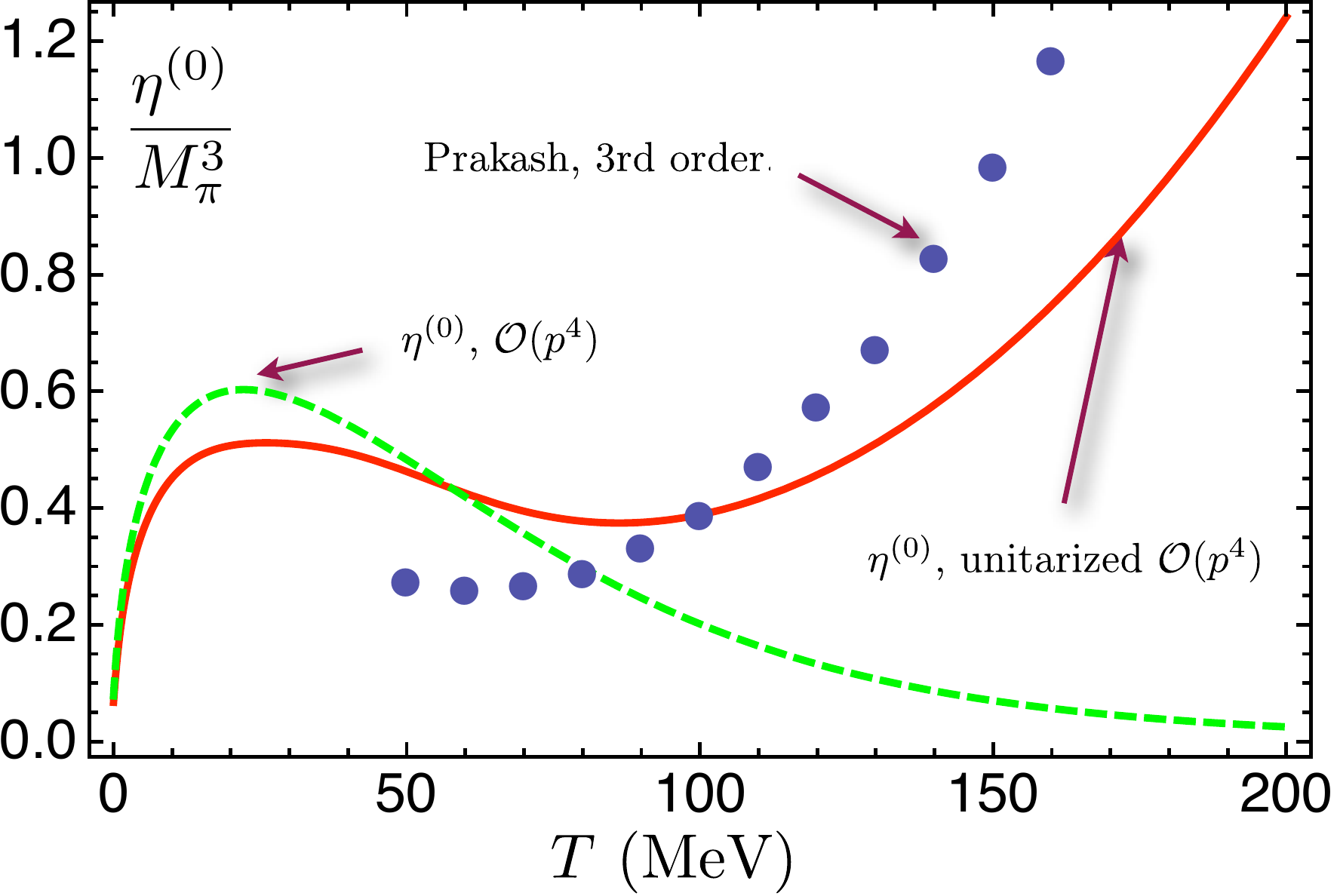}\\
\includegraphics[height=3.3cm]{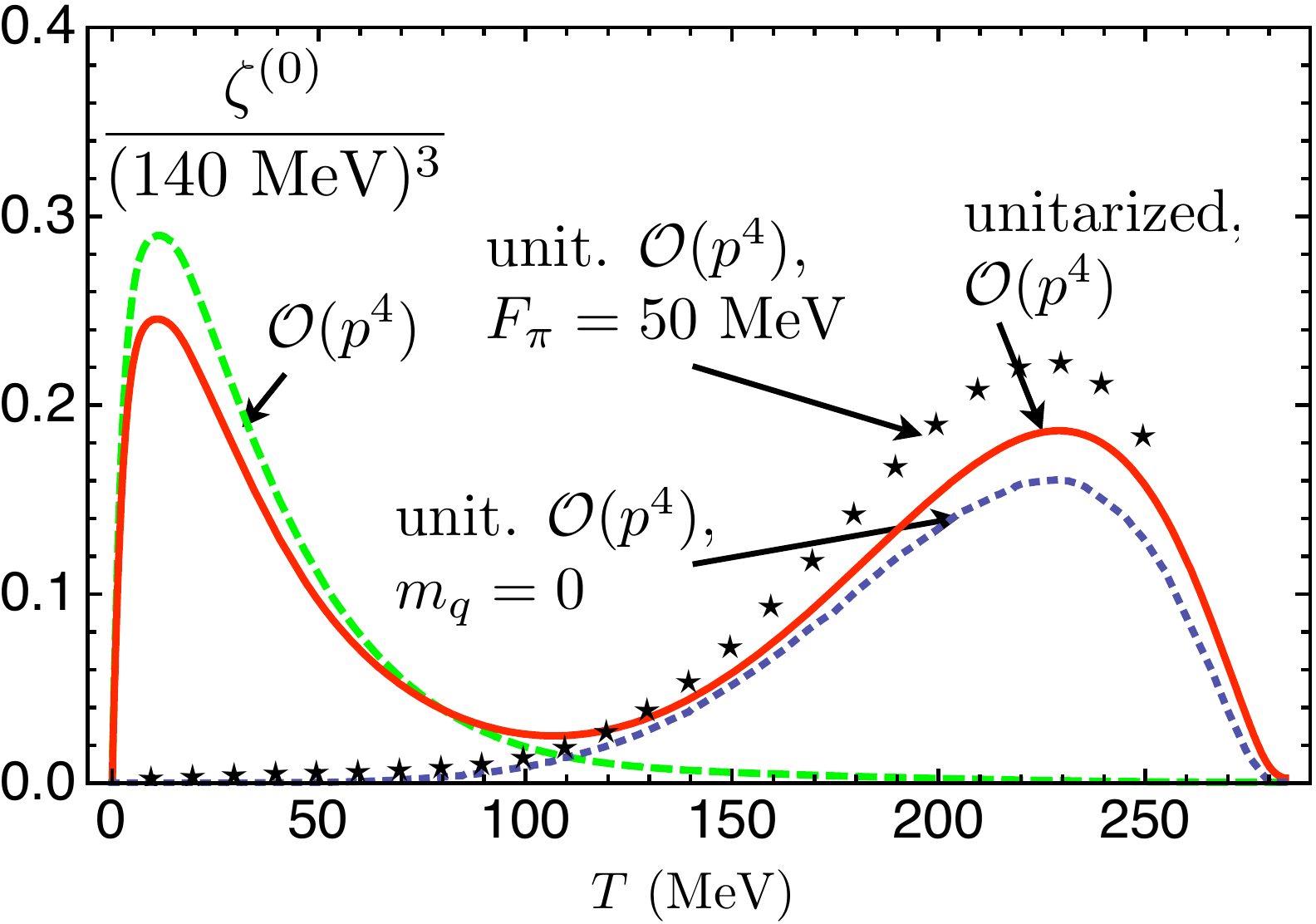}\hspace{0.5cm}\includegraphics[height=3.3cm]{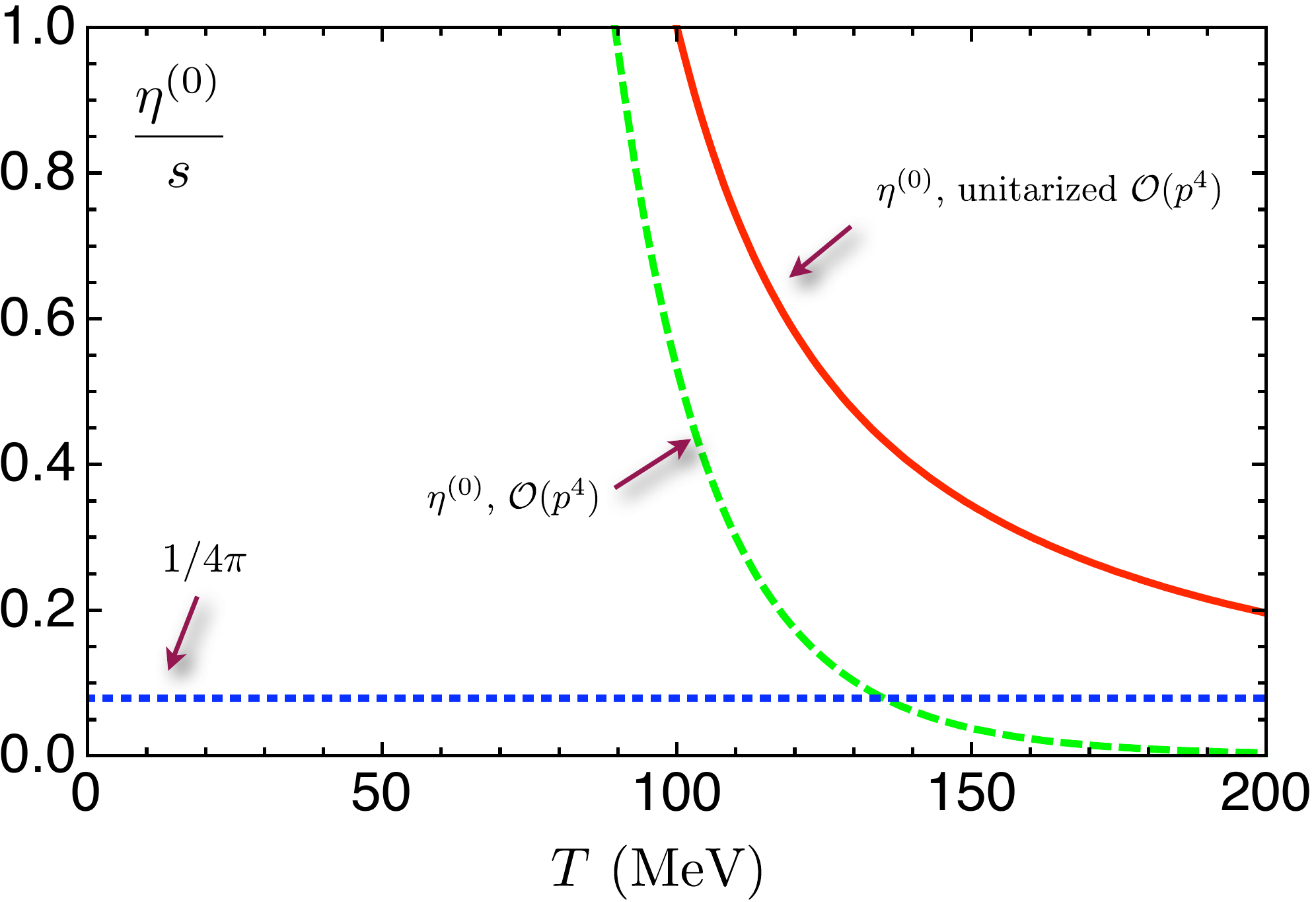}\hspace{0.5cm}\includegraphics[height=3.3cm]{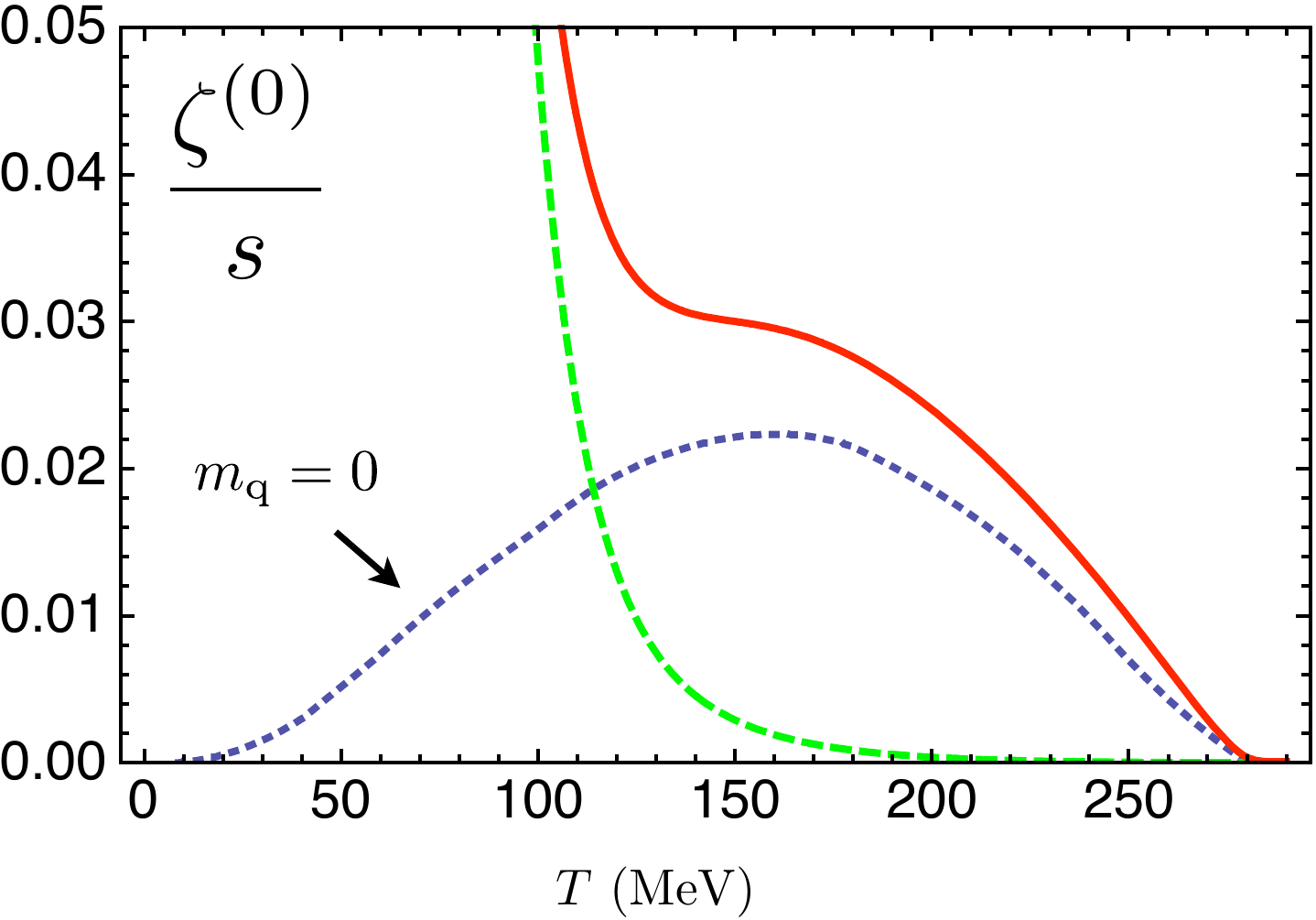}
\figcaption{\label{plotstransport} Leading order contribution (one
  loop) to the
  transport coefficients for a massive pion gas according to ChPT. The
solid red line represents the unitarized result to $\mathcal{O}(p^4)$
in the pion scattering amplitudes. The dashed green line represents
the non-unitarized result to $\mathcal{O}(p^4)$. For the thermal
conductivity and the shear viscosity we compare with the kinetic
theory results of [11].}
\end{center}
\ruledown
\begin{multicols}{2}

It has been recently proposed using a sum rule and an ansatz for the
spectral function that the bulk viscosity and the trace
anomaly in QCD should be correlated \cite{Kharzeev:07}, so that a maximum in the
trace anomaly would drive a maximum in $\zeta$. The spectral function
was modified very recently in \cite{Romatschke:09}, and the form of
the ansatz has been
questioned in \cite{CaronHuot:09}. So it is at the moment not clear how these two quantities are
actually related to each other. According to our results for the trace
anomaly in Fig. \ref{plottraceanom} and the bulk viscosity in
Fig. \ref{plotstransport}, we obtain a clear correlation between the
two \cite{FernandezFraile:09-2}, with a first peak associated to the
explicit breaking of the scale symmetry due to the finite quark mass,
and a second peak near $T_\mathrm{c}$ due to the scale anomaly.

In the near future we plan to analyze carefully the
massless case for the pion gas, vertex corrections in the calculation
of the bulk viscosity, and the inclusion of finite pion-number chemical
potential effects \cite{FernandezFraile:09-3}.

\begin{center}
\includegraphics[height=4cm]{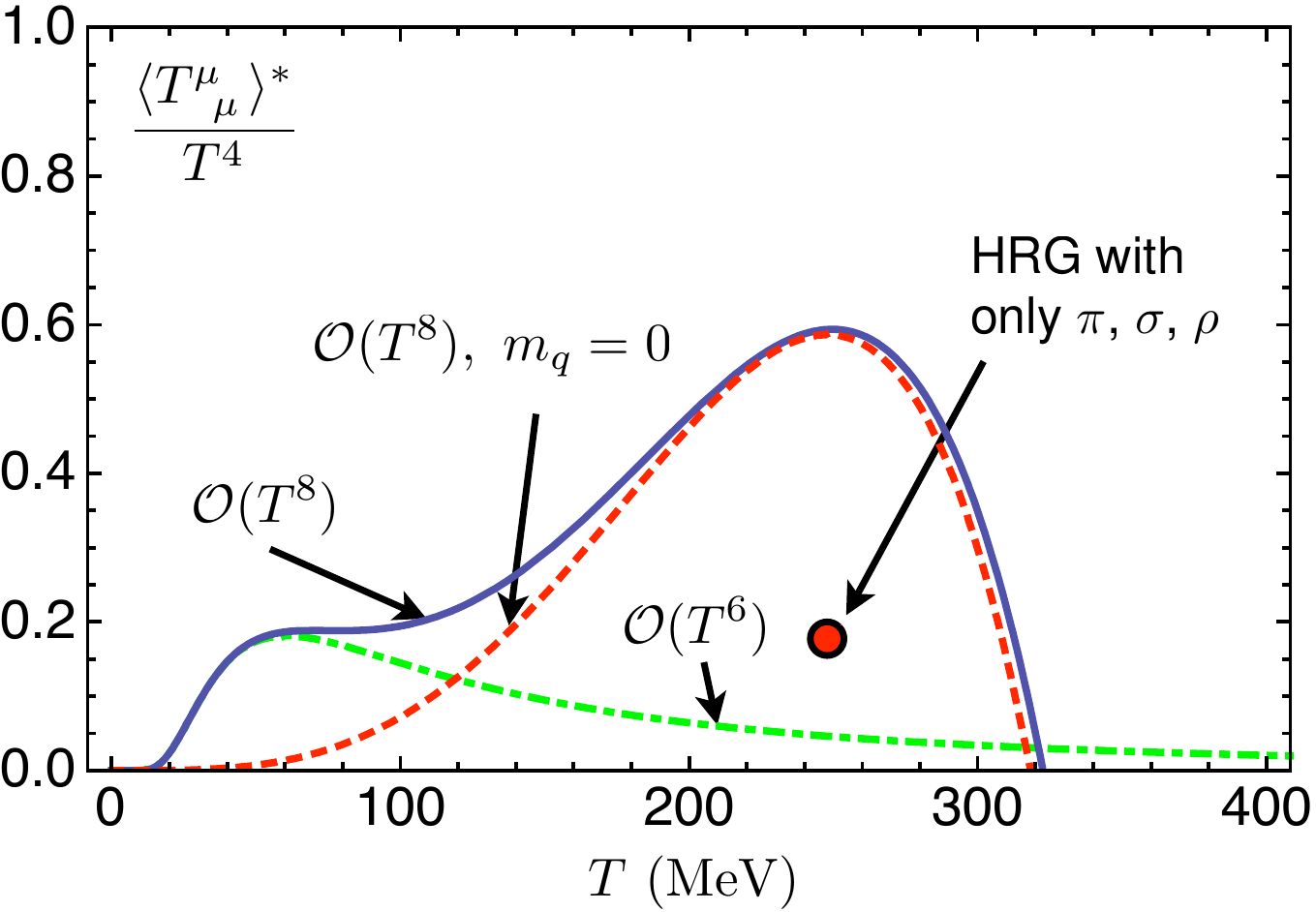}
\figcaption{\label{plottraceanom} Trace anomaly of the pion gas calculated in
  ChPT perturbatively.}
\end{center}

%%%%%%%%%%%%%%%%%%%%%%%%%%%%%%%%%%%%%%
\acknowledgments{The work of D.F.F. is sponsored by the
  Helmholtz International Center for FAIR. We both acknowledge
  financial support from the Spanish research Projects
  No. FPA2007-29115-E, No. PR34-1856-BSCH, No. UCM-BSCH GR58/08 910309,
  No. FPA2008-00592, No. FIS2008-01323.}
%%%%%%%%%%%%%%%%%%%%%%%%%%%%%%%%%%%%%%

\end{multicols}

\vspace{-2mm}
\centerline{\rule{80mm}{0.1pt}}
\vspace{2mm}

\begin{multicols}{2}

%%%%%%%%%%%%%%%%%%%%%%%%%%%%%%%%%%%%%

\end{multicols}
\clearpage
\end{document}